\documentclass{iaus}
\usepackage{graphicx}

\title[DLA Dwarf] 
{Connecting high-redshift galaxy populations through observations of local Damped Lyman Alpha dwarf galaxies}

\author[Regina Schulte-Ladbeck]   
{Regina E. Schulte-Ladbeck}

\affiliation{Department of Physics and Astronomy, University of Pittsburgh, \\ Pittsburgh, PA 15260, USA\\ 
email: {\tt rsl@pitt.edu}}

\pubyear{2008}
\volume{255}  
\pagerange{119--126}
\jname{Low-Metallicity Star Formation: From the First Stars to Dwarf Galaxies}
\editors{L.K. Hunt, S. Madden \& R. Schneider, eds.}
\begin{document}

\maketitle

\begin{abstract}
I report on observations of the z=0.01 dwarf galaxy
SBS1543+593 which is projected onto the background QSO HS1543+5921. As a star-forming
galaxy first noted in emission, this dwarf is playing a pivotal role in our understanding
of high-redshift galaxy populations, because it also gives rise to a Damped Lyman Alpha 
system. This enabled us to analyze, for the first time, the chemical abundance of
$\alpha$ elements in a Damped Lyman Alpha galaxy using both, emission and absorption diagnostics.
We find that the abundances agree with one another within the observational uncertainties. 
I discuss the implications of this result for the interpretation of high-redshift galaxy
observations. A catalog of dwarf-galaxy--QSO projections culled from the Sloan Digital Sky Survey is
provided to stimulate future work.
\end{abstract}

\firstsection 
\section{The Problem}

We now know a wealth of galaxy populations at high redshifts. Damped Lyman Alpha (DLA)
systems are identified from the strong neutral Hydrogen gas absorption they cause 
in background QSO spectra, while Lyman Break Galaxies, Lyman Alpha Emitters, and 
Gamma Ray Burst (GRB) host galaxies are discovered through emission from their stars 
and ionized gas. What we would yet like to understand is how these populations 
relate to each other, and to the galaxies we see today. While the study of DLA systems has made 
great contributions to our empirical knowledge of the chemical evolution of galaxies, unfortunately, 
despite decades of observational searches it has proven extremely difficult to detect 
the host galaxies directly in emission (cf. Dessauges-Zavadsky, this volume). Our understanding of high 
redshift galaxies remains biased by how they were selected, via absorption or emission. 

Figure~1 shows $\alpha$ element abundances 
for star-forming galaxies (SFG) and HII regions derived from emission lines (these are
O abundances), as well as for DLA systems determined from absorption lines (O, S, or Si abundances). 
This figure illustrates several key
points which have entered in the debate over the abundances of emission- versus absorption-selected 
systems. First, emission diagnostics tend to dominate the low-redshift 
abundance data while absorption diagnostics are still the method of choice for deriving high-redshift abundances.   
And second, there appears to be very little overlap in the $\alpha$ element abundances derived using
emission versus absorption techniques. What does that mean for the galaxy populations probed?

\section{Proposed Solutions}

Four solutions to this problem have been thought of at this time. They are briefly summarized below.

\noindent {\underline {\it Diagnostics Hypothesis}.} The diagnostics hypothesis posits that the absorption lines give lower 
metallicities than the emission lines because this is an inherent difference in the diagnostics. Verifying or
falsifying this hypothesis has been the focus of our work, and will be the main topic of this paper.

\noindent {\underline {\it Different Radial Biases Hypothesis}.} The different radial biases hypothesis predicts that lower abundances
will be measured in absorption than in emission if the negative radial abundance gradient observed in local spirals (\cite[Searle 1971]{Searle 1971})
holds for DLAs in general. Star-formation tends to occur in the centers of 
galaxies; QSO sightlines intercept foreground systems at any radii where the HI column density is above DLA treshold, thus, on
average, they will be probing larger radii. 

\noindent {\underline {\it Distinct Populations Hypothesis}.} The distinct population hypothesis postulates that 
low abundances for DLAs result if they are comprised of low-mass or dwarf (proto-)galaxies 
(\cite[Haehnelt et al.~1998]{Haehnelt1998}). Mass-metallicity relations have now been shown to exist out to 
redshifts of about 3 (cf. Mannucci, Lee, this volume). 

\noindent {\underline {\it Dust Obscuration Bias Hypothesis}.} A metal-rich interstellar medium tends to be dusty as well
(cf. Hunter, Spaans, this volume). The dust obscuration bias hypothesis predicts low abundances for DLAs result 
if optically selected QSO samples are biased toward QSOs with little foreground extinction (\cite[Ostriker \& Heisler 1984]{OH}).
 
\section{Testing the Diagnostics Hypothesis}
The emission-line technique is based on the measurement of forbidden-line fluxes. Line fluxes 
need to be corrected for stellar and foreground dust absorptions. 
We then determine electron temperature, density, and ionic abundances, which, with appropriate
ionization corrections, lead to element abundances. When the temperature cannot be derived, and for high-redshift galaxies, we often 
use a strong-line method to evaluate abundances rather than the direct or Te method.
The emission-line technique for deriving abundances is described more fully in Stasinska (this volume).

The absorption-line technique is based on the measurement of absorption-line profiles. Their optical depths lead to ionic column densities. 
Dust depletion and ionization corrections must be considered; then we determine element abundances. 
When high-resolution spectra are not available, such as is the case with
FUSE and HST data of DLAs, we use the curve-of-growth method to estimate an ion's column density.

What do we know about whether the physics of the line diagnostics works for astronomical objects? 
\cite[Williams et al. (2008)]{W08} have recently compared the abundances of planetary nebulae measured in emission
with the abundances measured in absorption toward their central stars. They find good agreement between forbidden emission-line and absorption-line
abundances for a wide range of elements.

What do we know about how the diagnostics work on the scale of galaxies? And wouldn't the abundances depend on whether and how well 
the different phases of the interstellar medium in which they are measured are mixed? In the Galaxy, HII regions abundances interpolated 
for the distance of the solar circle (\cite[Deharveng et al. 2000]{D0})
agree very well with abundances in the local ISM measured on sightlines toward nearby hot stars (\cite[Moos et al. 2002]{M02}). 
This leads to the expectation that 
emission and absorption diagnostics should also give concordant abundances when applied to other galaxies. 

\begin{figure}[b]
\vspace*{-1.0 cm}
\begin{center}
 \includegraphics[width=3.4in]{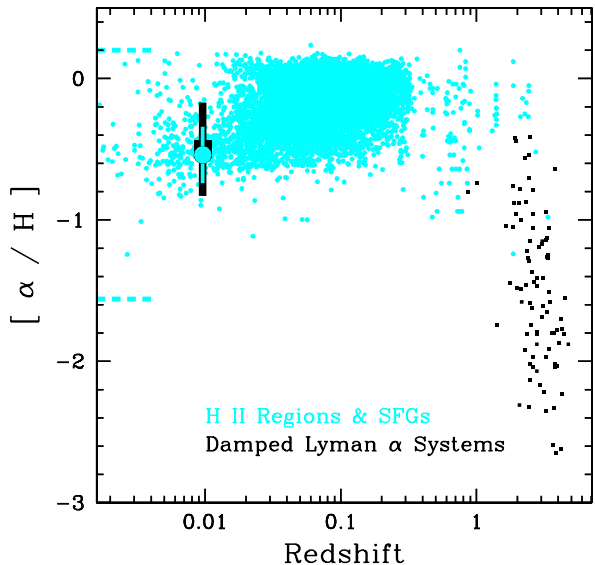} 
 \caption{Alpha element abundances for SFGs and DLAs (adapted from \cite[K{\"o}nig et al.~2006]{Konig2006}). The dashed lines indicate the 
 highest and lowest metallicities measured in the local galaxy population. Our measurements for the
 [O/H]$_{II}$ and [S/H]$_{I}$ abundances of SBS1543+593 are shown as well.}
   \label{fig1}
\end{center}
\end{figure}
\subsection{Test Designs}

There are two experimental designs that have been applied to the study of emission- and absorption-based abundances in external galaxies. 
The first method uses an internal
starcluster or starburst within a galaxy (cf. contributions by Lebouteiller, Aloisi, and Thuan in this volume). The advantage of doing so is that the
HII and the HI gas are both on the sightline to the starcluster. The disadvantage is that the starcluster 
may contain thousands of hot stars and is spatially extended; it also lies behind a complex foreground screen. 
The size of the aperture used for the absorption spectroscopy usually results in an integration
over these many, diverse sightlines. This is a potential problem that could be fixed if a single hot star could be picked out from the cluster, 
but that has not been possible due to the high spatial resolution that would be required.

The second method employs the classical absorption technique by using a QSO as the background source. The advantage is that the background source is
clearly a point source. The disadvantage is that the probability of finding a QSO sightline that is aligned with an HII region in a foreground galaxy is
extremely small. Recall it has been hard to find DLA galaxies at all. We therefore cannot, at this time, measure the HII abundances
on the same sightlines as the HI abundances. This could be fixed if we could find a QSO behind an HII region. A possible workaround is to 
use an HII region in a dwarf galaxy, because, to the best of our knowledge, dwarf galaxies do not exhibit radial abundance gradients 
(cf. Stasinska, this volume). 
For this to work, one has to find suitable dwarf DLA galaxy--QSO projections. Until recently, there was only one known case.

\subsection{SBS1543+593}

The alignment of SBS1543+593 (z=0.0096) with the QSO HS1543+5921 was discovered by \cite[Reimers \& Hagen (1998)]{reimers1998}. 
Previously, the galaxy had been thought of as having a Seyfert
nucleus. The discovery that the galaxy intercepts the QSO at a mere 0.5~kpc finally enabled the test of diagnostics hypothesis. 
\cite[Bowen et al. (2001)]{bowen2001} found that the
sightline to the QSO gives rise to a DLA. \cite[Schulte-Ladbeck et al. (2004)]{SL4} classified the galaxy an Sdm, 
determined its absolute B magnitude (-16.8) and estimated its 
star-formation rate (0.006 M$_{\odot}$yr$^{-1}$). We also derived 12+log(O/H)$_{II}$ is 8.2$\pm$0.2, or [O/H]$_{II}$ is -0.54$\pm$0.2. In 
\cite[Schulte-Ladbeck et al. (2005)]{SL5} we also
derive a Sulphur abundance in emission, and determine [S/H]$_{II}$ is -0.27$\pm$0.3. Our analysis of HST spectra, on the other hand, yielded 
[O/H]$_{I}$$>$-2.14 (the line tends to be saturated) and [S/H]$_{I}$=-0.50$\pm$0.33 (this may be a lower limit as the lines were unresolved
and may therefore contain hidden saturated components).
The [O/H]$_{II}$ and [S/H]$_{I}$ data are poltted in Fig.~1.

\subsection{Result}
Our experiment indicates that the emission- and absorption-derived $\alpha$ element abundances of SBS1543+593 agree within the errors.
The {\bf diagnostics hypothesis} has been {\bf falsified}. In other words, the hypothesis that QSO absorption lines give lower
abundances than HII emission lines for (proto-)galactic systems is not true.

\subsection{Verification}
Analysis of the same HST data by \cite[Bowen et al. (2005)]{B5} resulted in [O/H]$_{I}$$>$-0.9 and [S/H]$_{I}$=-0.41$\pm$0.06, 
in good agreement with Schulte-Ladbeck et al. (2004, 2005).
There has been no independent confirmation of the ionized gas abundances in SBS1543+593.

\subsection{Implications}
Our result indicates that abundances derived via HII-region emission lines and QSO absorption lines are directly comparable. The difference
between the abundances measured in SFGs and DLAs is real. Our result also implies that the neutral gas of SBS1543+593, which extends 
out to 15 times its optical radius 
(\cite[Rosenberg et al. 2006]{R6}), 
has the same abundance as the ionized gas found within the confines of the galaxy. 

Abundances for high-redshift GRB host galaxies are now routinely determined from absorption lines. This is analogous to the QSO experiment since the GRB
is a point source. Swift recently discovered a wealth of low-redshift GRBs. With HST lacking UV spectroscopic ability, the abundances of the host 
galaxies have been detemined using emission lines. Savaglio (this volume) shows an abundance-redshift plot for GRB host galaxies, 
combining the low-redshift emission abundances with the high-redshift absorption abundances, to discuss trends in
metallicity evolution of GRBs. Our result indicates that this is indeed a valid comparison.

\section{Future Work: A Catalog of Dwarf Galaxies on QSO Sightlines}

It would be helpful to obtain confirmation of our result through observation of more dwarf galaxies on QSO sightlines. 
Table~1 is a catalog of SDSS dwarf galaxy (M$_B$$>$-18) \& QSO pairs from B. Cherinka's thesis project. It gives the SDSS
galaxy, QSO name, redshifts, galaxy luminosity (k- and foreground absorption corrected, in the g band), 
and ratio of impact parameter to galaxy r-band radius. Based on the strength of their CaII or NaI QSO absorption lines, potential candidates for
DLA galaxies are SDSS J032803.11+002055.1, J125700.31+010143.3 (=UGC~8066), J170330.32+240330.8, and J221216.89+003243.9. 
It should be noted that the SDSS QSO spectra are of quite low S/N, therefore, our absorption line measurements are quite uncertain. UGC~8066, an LSB
galaxy with log~M$_{HI}$[M$_{\odot}$]=9.29 and M$_{HI}$/L$_B$$\approx$2 (\cite[Burkholder et al. 2001]{B1}), is perhaps our best candidate. 

\begin{table}
 \begin{center}
 \caption{Dwarf Galaxies on QSO sightlines in SDSS} \label{tab1}
 {\scriptsize
 \begin{tabular}{|l|c|c|c|c|c|}\hline
{\bf SDSS Galaxy Name} & {$\bf z_{GAL}$} & {\bf SDSS QSO Name} & {$\bf
z_{QSO}$} & {$\bf L_g^{k,f}$} & {$\bf b/r_{petro}$}\\
  &  & &  & {$\bf [L_g^*]$} &  \\ \hline
J001233.41+010014.2  & 0.08543 & J001233.34+010010.3 & 1.21324& 0.16 & 1.11\\ \hline
J021734.23-002637.2  & 0.04069 & J021734.63-002641.9 & 1.55744& 0.10 & 1.33 \\ \hline
J023818.88-003030.5  & 0.03724 & J023819.26-003029.3 & 2.60503& 0.14 & 0.85 \\ \hline
J024329.07+003833.7  & 0.02796 & J024328.86+003831.2 & 2.75318& 0.07 & 0.59 \\ \hline
J024421.09+004031.3  & 0.00932 & J024420.36+004029.2 & 2.21791& 0.09 & 0.42 \\ \hline
J032758.83+001652.0  & 0.03702 & J032759.51+001713.1 & 2.06170& 0.04 & 0.84 \\ \hline
J032803.11+002055.1  & 0.02363 & J032801.70+002100.1 & 0.32205& 0.02 & 1.91 \\ \hline
J075010.54+304106.3  & 0.01477 & J075010.17+304032.3 & 1.89210& 0.07 & 1.82 \\ \hline
J112023.20+574429.5  & 0.00697 & J112020.12+340555.3 & 0.76961& 0.01 & 1.30 \\ \hline
J113955.50+132802.0  & 0.01193 & J113955.97+132713.3 & 1.99390& 0.09 & 1.12 \\ \hline
J115115.25+485331.0  & 0.02564 & J115118.58+485331.1 & 1.07180& 0.11 & 1.18 \\ \hline
J122754.83+080525.4  & 0.00207 & J122752.60+080526.6 & 1.62126& 0.01 & 1.94 \\ \hline
J123636.73+141333.1  & 0.00374 & J123637.35+141316.2 & 1.60133& 0.002 & 1.37 \\ \hline
J125700.31+010143.3  & 0.00930 & J125703.67+010132.0 & 0.958967& 0.04 & 1.18 \\ \hline
J131529.74+472958.7  & 0.00086 & J131531.57+473054.6 & 1.72172& 0.001 & 1.69 \\ \hline
J170330.32+240330.8  & 0.03083 & J170331.83+240339.8 & 0.95816& 0.16 & 1.84 \\ \hline
J221216.89+003243.9  & 0.02971 & J221217.27+003227.0 & 2.25410& 0.08 & 0.99 \\ \hline
J233724.00+002330.0  & 0.00932 & J233722.01+002238.9 & 1.37617& 0.07 & 0.87 \\ \hline
 \end{tabular}
 }
 \end{center}
\vspace{1mm}
 \scriptsize{
}
\end{table}

\bigskip
\noindent {\bf {Acknowledgements.}} I thank my department chair, David Turnshek, for approving a travel grant that helped 
offset some of the cost of my conference participation. 
Brian Cherinka and I acknowledge the use of SDSS data (see sdss.org/collaboration/credits.html).


\begin{thebibliography}{}

\bibitem[Bowen et al. 2001]{bowen2001} Bowen, D.~V.,
 Tripp, T.~M., \& Jenkins, E.~B.\ 2001, \textit{AJ}, 121, 1456
 
\bibitem[Bowen et al. 2005]{B5} Bowen, D.~V., Jenkins, 
E.~B., Pettini, M., \& Tripp, T.~M.\ 2005, \textit{ApJ}, 635, 880 

\bibitem[Burkholder et al. 2001]{B1} Burkholder, V., 
Impey, C., \& Sprayberry, D.\ 2001, \textit{AJ}, 122, 2318 
 
\bibitem[Deharveng et al. 2000]{D0} Deharveng, L., 
Pe{\~n}a, M., Caplan, J., \& Costero, R.\ 2000, \textit{MNRAS}, 311, 329 

\bibitem[K{\"o}nig et al. 2006]{Konig2006} 
{K{\"o}nig, B., Schulte-Ladbeck, R.~E., \& Cherinka, B.} 2006, 
\textit{AJ}, 132, 1844 

\bibitem[Haehnelt et al.~1998]{Haehnelt1998}
{Haehnelt, M.~G., Steinmetz, M., \& Rauch, M.} 1998,
\textit{ApJ}, 495, 647

\bibitem[Moos et al. 2002 ]{M02} Moos, H.~W., et al.\ 2002, 
\textit{ApJS}, 140, 3 

\bibitem[Ostriker \& Heisler 1984]{OH}
{Ostriker, J.~P., \& Heisler, J.} 1984,
\textit{ApJ}, 278, 1

\bibitem[Reimers \& Hagen 1998]{reimers1998} Reimers, D.~\& Hagen,
  H.-J.\ 1998, \textit{A\&A}, 329, L25
  
\bibitem[Rosenberg et al. 2006]{R6} Rosenberg, J.~L., 
Bowen, D.~V., Tripp, T.~M., \& Brinks, E.\ 2006, \textit{AJ}, 132, 478   

\bibitem[Schulte-Ladbeck et al. 2004]{SL4} Schulte-Ladbeck, R.~E., Rao, S.~M., Drozdovsky, I.~O., Turnshek, D.~A., 
Nestor, D.~B., \& Pettini, M.\ 2004, \textit{ApJ}, 600, 613 
  
\bibitem[Schulte-Ladbeck et al. 2005]{SL5} 
Schulte-Ladbeck, R.~E., K{\"o}nig, B., Miller, C.~J., Hopkins, A.~M., 
Drozdovsky, I.~O., Turnshek, D.~A., \& Hopp, U.\ 2005, \textit{ApJL}, 625, L79 
   
\bibitem[Searle 1971]{Searle 1971}
{Searle, L.} 1971
\textit{ApJ}, 168, 327

\bibitem[Williams et al.2008]{W08}
{Williams, R., Jenkins, E.~B., Baldwin, J.~A., Zhang, Y., Sharpee, B., Pellegrini, E., 
\& Phillips, M.} 2008,
\textit{ApJ}, 677, 1100

\end{thebibliography}
\end{document}